\PassOptionsToPackage{unicode}{hyperref}
\PassOptionsToPackage{hyphens}{url}
\documentclass[
]{article}
\usepackage{amsmath,amssymb}
\usepackage{iftex}
\ifPDFTeX
  \usepackage[T1]{fontenc}
  \usepackage[utf8]{inputenc}
  \usepackage{textcomp} 
\else 
  \usepackage{unicode-math} 
  \defaultfontfeatures{Scale=MatchLowercase}
  \defaultfontfeatures[\rmfamily]{Ligatures=TeX,Scale=1}
\fi
\usepackage{lmodern}
\ifPDFTeX\else
\fi
\IfFileExists{upquote.sty}{\usepackage{upquote}}{}
\IfFileExists{microtype.sty}{
  \usepackage[]{microtype}
  \UseMicrotypeSet[protrusion]{basicmath} 
}{}
\makeatletter
\@ifundefined{KOMAClassName}{
  \IfFileExists{parskip.sty}{%
    \usepackage{parskip}
  }{
    \setlength{\parindent}{0pt}
    \setlength{\parskip}{6pt plus 2pt minus 1pt}}
}{
  \KOMAoptions{parskip=half}}
\makeatother
\usepackage{xcolor}
\usepackage[margin=1in]{geometry}
\usepackage{color}
\usepackage{fancyvrb}

\DefineVerbatimEnvironment{Highlighting}{Verbatim}{commandchars=\\\{\}}
\usepackage{framed}
\definecolor{shadecolor}{RGB}{248,248,248}
\newenvironment{Shaded}{\begin{snugshade}}{\end{snugshade}}

\newcommand{\AttributeTok}[1]{\textcolor[rgb]{0.13,0.29,0.53}{#1}}

\newcommand{\CommentTok}[1]{\textcolor[rgb]{0.56,0.35,0.01}{\textit{#1}}}

\newcommand{\ConstantTok}[1]{\textcolor[rgb]{0.56,0.35,0.01}{#1}}

\newcommand{\DecValTok}[1]{\textcolor[rgb]{0.00,0.00,0.81}{#1}}

\newcommand{\FloatTok}[1]{\textcolor[rgb]{0.00,0.00,0.81}{#1}}
\newcommand{\FunctionTok}[1]{\textcolor[rgb]{0.13,0.29,0.53}{\textbf{#1}}}

\newcommand{\NormalTok}[1]{#1}

\newcommand{\OtherTok}[1]{\textcolor[rgb]{0.56,0.35,0.01}{#1}}

\newcommand{\SpecialCharTok}[1]{\textcolor[rgb]{0.81,0.36,0.00}{\textbf{#1}}}

\newcommand{\StringTok}[1]{\textcolor[rgb]{0.31,0.60,0.02}{#1}}

\usepackage{longtable,booktabs,array}
\usepackage{calc} 
\usepackage{etoolbox}
\makeatletter
\patchcmd\longtable{\par}{\if@noskipsec\mbox{}\fi\par}{}{}
\makeatother
\IfFileExists{footnotehyper.sty}{\usepackage{footnotehyper}}{\usepackage{footnote}}
\makesavenoteenv{longtable}
\usepackage{graphicx}
\makeatletter
\def\maxwidth{\ifdim\Gin@nat@width>\linewidth\linewidth\else\Gin@nat@width\fi}
\def\maxheight{\ifdim\Gin@nat@height>\textheight\textheight\else\Gin@nat@height\fi}
\makeatother
\setkeys{Gin}{width=\maxwidth,height=\maxheight,keepaspectratio}
\makeatletter
\def\fps@figure{htbp}
\makeatother
\providecommand{\tightlist}{%
  \setlength{\itemsep}{0pt}\setlength{\parskip}{0pt}}
\NewDocumentCommand\citeproctext{}{}

\makeatletter
 \let\@cite@ofmt\@firstofone
 \def\@biblabel#1{}
 \def\@cite#1#2{{#1\if@tempswa , #2\fi}}
\makeatother
\newlength{\cslhangindent}
\newlength{\csllabelwidth}
\newenvironment{CSLReferences}[2] 
 {\begin{list}{}{%
  \setlength{\itemindent}{0pt}
  \setlength{\leftmargin}{0pt}
  \setlength{\parsep}{0pt}
  \ifodd #1
   \setlength{\leftmargin}{\cslhangindent}
   \setlength{\itemindent}{-1\cslhangindent}
  \fi
  \setlength{\itemsep}{#2\baselineskip}}}
 {\end{list}}
\usepackage{calc}

\ifLuaTeX
  \usepackage{selnolig}  
\fi
\usepackage{bookmark}
\IfFileExists{xurl.sty}{\usepackage{xurl}}{} 
\hypersetup{
  pdftitle={An introduction to R package mvs},
  pdfauthor={Wouter van Loon},
  hidelinks,
  pdfcreator={LaTeX via pandoc}}

\title{An introduction to R package \texttt{mvs}}
\author{Wouter van Loon}
\date{April 15, 2025}

\begin{document}
\maketitle

\section{Package summary}\label{package-summary}

In biomedical science, a set of objects or persons can often be described by multiple distinct sets of \emph{features} (also called \emph{independent variables} or \emph{predictors}) obtained from different data sources or modalities (Y. Li, Wu, and Ngom 2018). These feature sets are said to provide different \emph{views} of the objects or persons under consideration. Data sets consisting of multiple views are called \emph{multi-view data} (J. Zhao et al. 2017; Sun et al. 2019; Smilde, Næs, and Liland 2022)\footnote{Depending on the research area, multi-view data is sometimes called multi-block, multi-set, multi-group, or multi-table data (Smilde, Næs, and Liland 2022).}. Classical machine learning methods ignore the multi-view structure of such data, limiting model interpretability and performance. The R package \texttt{mvs} provides methods that were designed specifically for dealing with multi-view data, based on the \emph{multi-view stacking} (MVS) framework (R. Li et al. 2011; Garcia-Ceja, Galván-Tejada, and Brena 2018; Van Loon et al. 2020). MVS is a form of supervised\footnote{Like classical machine learning, multi-view learning can be divided into supervised, unsupervised, and semi-supervised learning. Supervised learning means there is an outcome variable that is used to guide the learning process (Friedman, Hastie, and Tibshirani 2009). Unsupervised learning means that only the features are observed and there is no known outcome variable (Friedman, Hastie, and Tibshirani 2009). Semi-supervised learning refers to a setting where there is an outcome variable of interest, but it has only been measured for a subset of the observations (Goodfellow, Bengio, and Courville 2016). The methods provided by the R package \texttt{mvs} are supervised learning methods. However, they could also be used for semi-supervised learning by treating the unobserved values of the outcome variable as missing data and using the provided imputation methods (see \hyperref[handling-missing-data]{Handling missing data}). For an overview of multi-view learning methods specific to semi-supervised or unsupervised settings see, for example, Sun et al. (2019) and Smilde et al. (2022).} (machine) learning used to train multi-view classification or prediction models. MVS works by training a learning algorithm (the \emph{base-learner}) on each view separately, estimating the predictive power of each view-specific model through cross-validation, and then using another learning algorithm (the \emph{meta-learner}) to assign weights to the view-specific models based on their estimated predictions. MVS is a form of \emph{ensemble learning}, dividing the large multi-view learning problem into smaller sub-problems. Most of these sub-problems can be solved in parallel, making it computationally attractive. Additionally, the number of features of the sub-problems is greatly reduced compared with the full multi-view learning problem. This makes MVS especially useful when the total number of features is larger than the number of observations (i.e., \emph{high-dimensional} data). MVS can still be applied even if the sub-problems are themselves high-dimensional by adding suitable \emph{penalty terms} to the learning algorithms (Van Loon et al. 2020). Furthermore, MVS can be used to automatically select the views which are most important for prediction (Van Loon et al. 2020). The R package \texttt{mvs} makes fitting MVS models, including such penalty terms, easily and openly accessible. \texttt{mvs} allows for the fitting of stacked models with any number of levels, with different penalty terms, different outcome distributions, and provides several options for missing data handling.

\section{What is multi-view stacking?}\label{what-is-multi-view-stacking}

Consider the following hypothetical example: We want to build a model that can predict whether or not a person has Alzheimer's disease based on different sources of data. Additionally, we want to find out which sources are most predictive of the outcome. That way, if certain sources turn out not to be predictive, they can be omitted from future data collection. We have collected data from 200 research subjects, half of which have been diagnosed with Alzheimer, and half are healthy controls. For each subject, we have the following sources of data available:

\begin{enumerate}
\def\labelenumi{\arabic{enumi}.}
\tightlist
\item
  A structural magnetic resonance imaging (MRI) scan, containing information about the structure of the brain.
\item
  A resting-state functional MRI (fMRI) scan, containing information about the functioning of the brain at rest.
\item
  A blood or cerebrospinal fluid (CSF) sample, from which genetic information can be derived.
\end{enumerate}

Now we could just collect all the features (independent variables) from these three \emph{views} together into a single data frame or matrix and fit a `traditional' feature-selecting model on the complete data. Combining the views together like this is called \emph{feature concatenation}. However, this approach typically ignores the multi-view structure of the data. To take the multi-view structure into account, we can instead fit a multi-view stacking (MVS) model. Multi-view stacking has several advantages over feature concatenation:

\begin{enumerate}
\def\labelenumi{\arabic{enumi}.}
\tightlist
\item
  It explicitly takes into account the multi-view structure of the data.
\item
  It divides a big model training problem into smaller sub-problems, which are easier to compute and can be computed in parallel.
\item
  It estimates a regression coefficient \emph{for each view} as well as for each feature.
\item
  The regression coefficients of the views are based on estimated out-of-sample predictions, improving generalization.
\end{enumerate}

\begin{figure}

{\centering \includegraphics[width=0.7\linewidth]{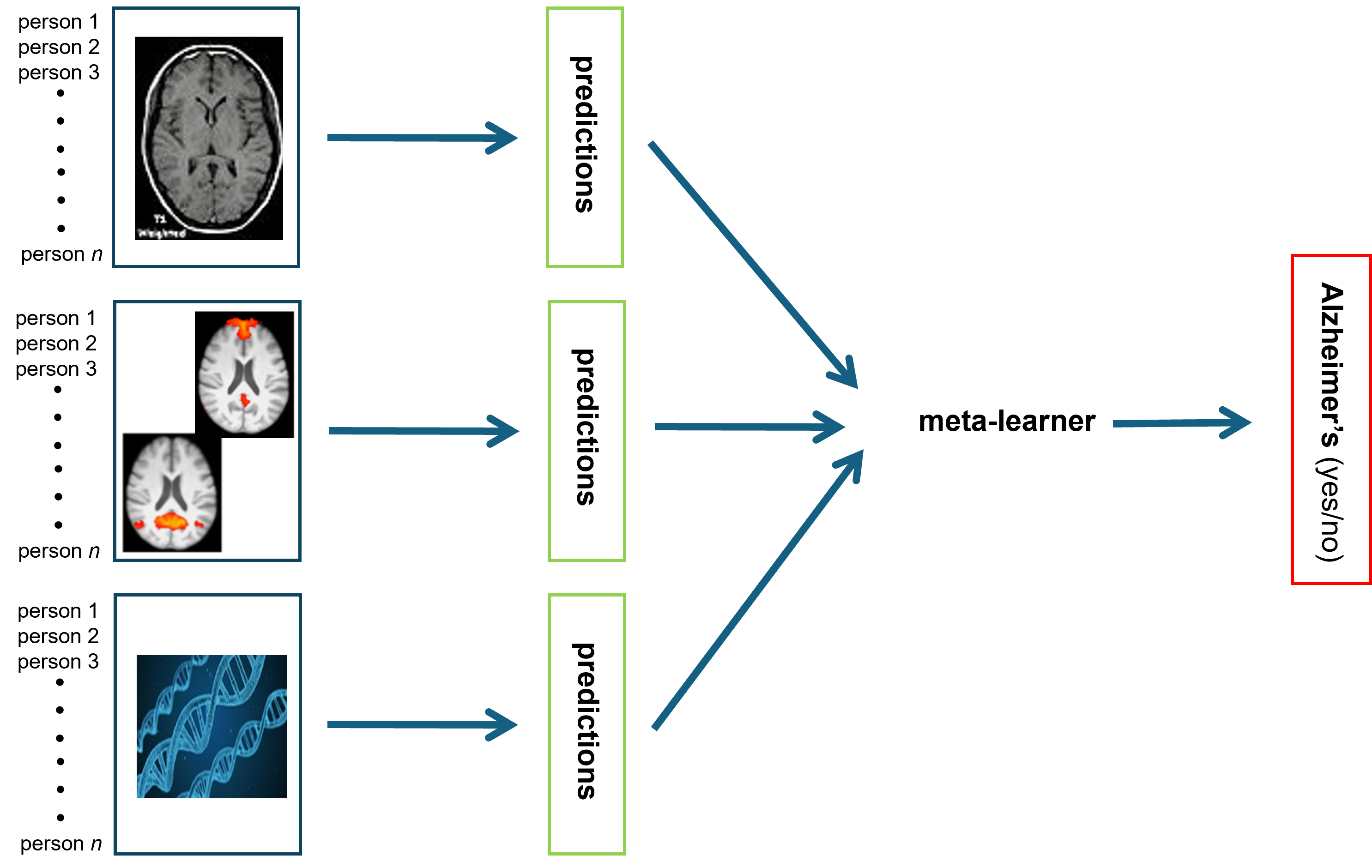} 

}

\caption{A simple graphic representation of a multi-view stacking model including 3 views: structural MRI, functional MRI, and genetic information. A sub-model is fitted on each view separately, and the predictions of these sub-models are combined by the meta-learner into a single prediction. Note that the *n* persons are the same persons for each view.}\label{fig:mvs-diag}
\end{figure}

A multi-view stacking model fitted on this data could look something like Figure \ref{fig:mvs-diag}.

\begin{figure}

{\centering \includegraphics[width=0.7\linewidth]{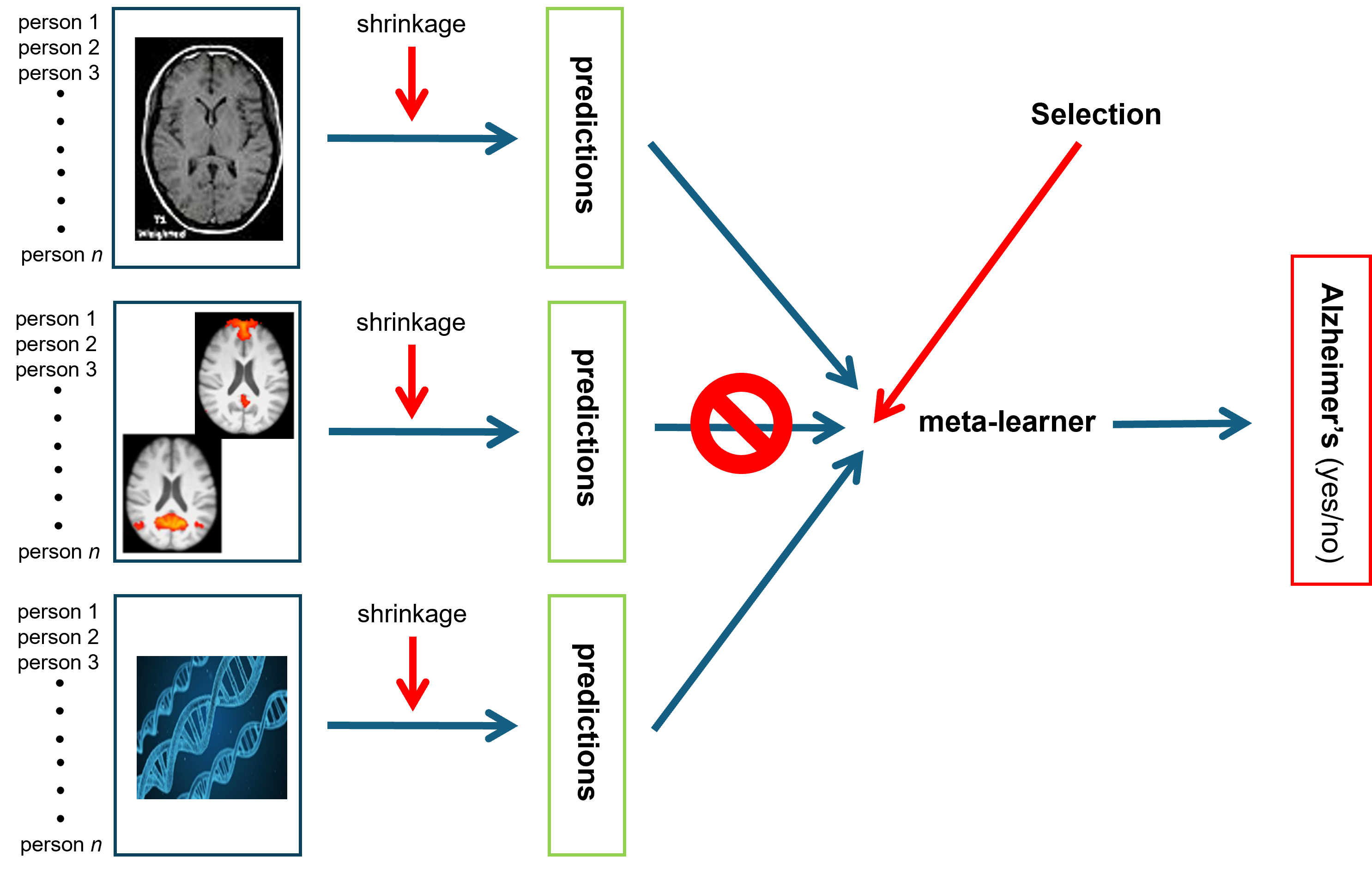} 

}

\caption{A simple graphic representation of how StaPLR can perform automatic view selection. In this (hypothetical) example, functional MRI was discarded from the model because it was not sufficiently predictive of the outcome in the presence of the other two views.}\label{fig:staplr-diag}
\end{figure}

Multi-view stacking is a very broad framework: Any suitable learning algorithm can be chosen as the base and meta-learner. In this case, the outcome variable is binary (yes/no), so we will want to use some sort of classifier. Additionally, suitable penalty terms could be added to these classifiers to automatically select the views that are most important for prediction. Stacked penalized logistic regression (StaPLR) is a form of multi-view stacking where we use penalized logistic regression as both the base and meta-learner. If we use a penalty term that induces selection (such as the \emph{lasso} (Tibshirani 1996)) in the meta-learner, it will automatically select the most important views. We also typically put additional nonnegativity constraints on the meta-learner, for both technical reasons and to improve interpretability (Van Loon et al. 2020). Since the views most likely contain many features (data obtained from fMRI scans can contain millions of features (Van Loon et al. 2022)), we may apply a penalty term in the base-learner that induces shrinkage (such as \emph{ridge regression}). If we apply StaPLR to the data, we may find that one of the views is not predictive of the outcome, like in Figure \ref{fig:staplr-diag}.

Of course, Figures \ref{fig:mvs-diag} and \ref{fig:staplr-diag} are simplified and represent only the trained model. The training process itself is slightly more complicated, due to the inclusion of a cross-validation step. A more technically accurate description of MVS is given in Algorithm 1.

\begin{figure}

{\centering \includegraphics[width=0.7\linewidth]{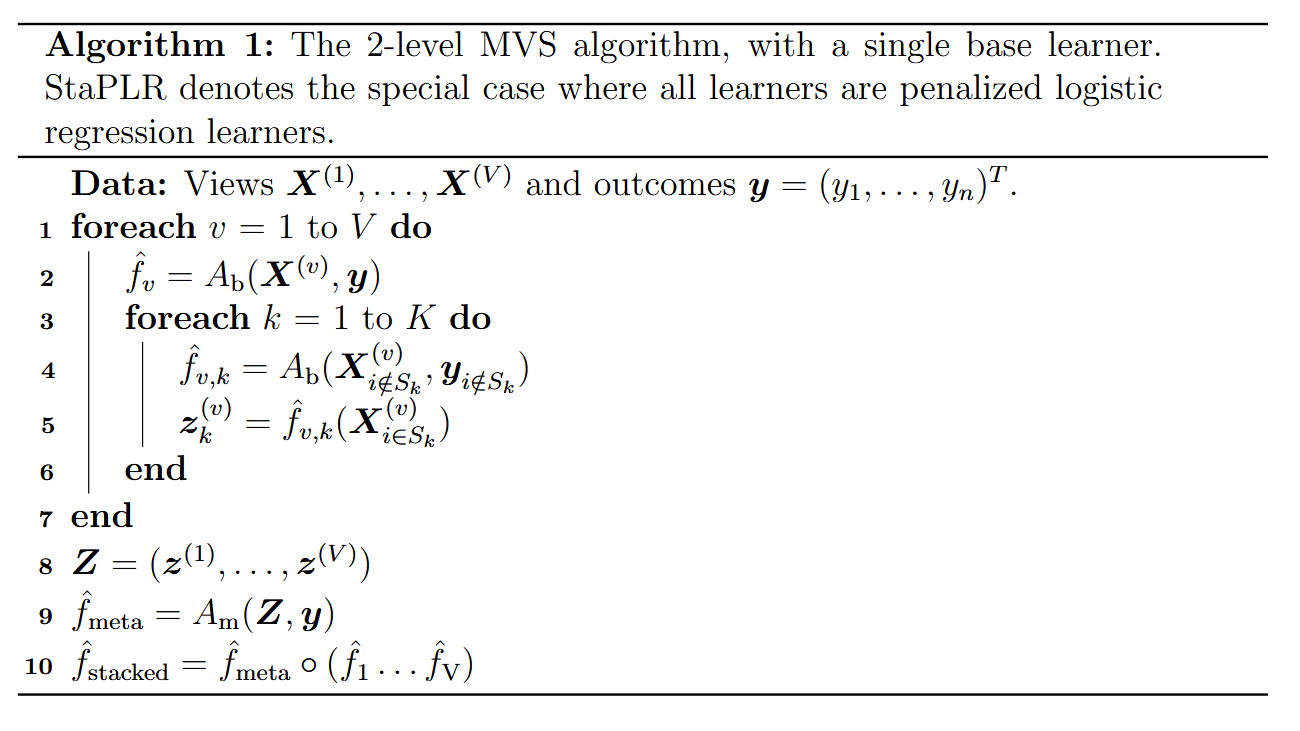} 

}

\end{figure}

\begin{figure}

{\centering \includegraphics[width=0.7\linewidth]{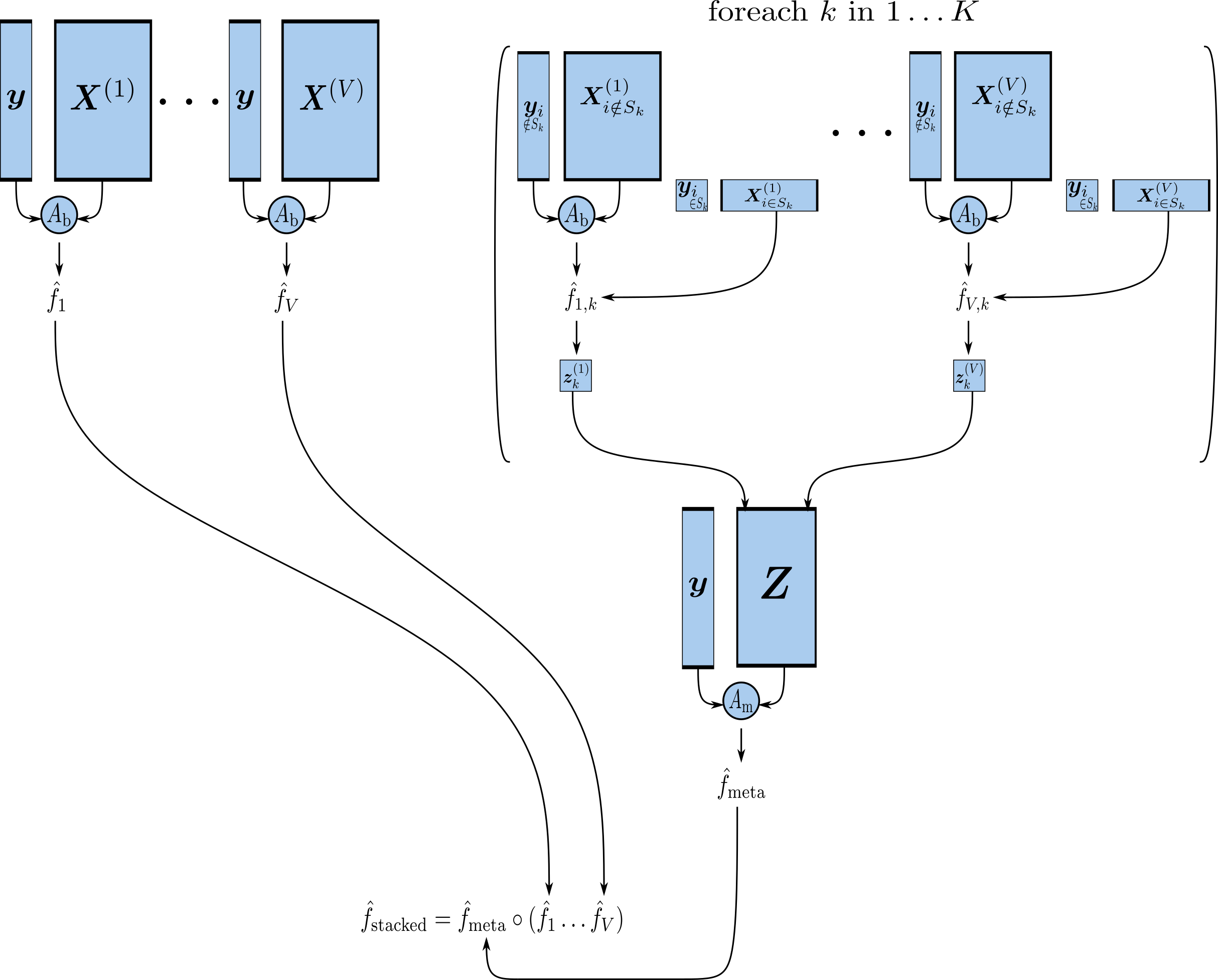} 

}

\caption{The MVS algorithm represented as a flow diagram. StaPLR denotes the special case where all learners are penalized logistic regression learners. Figure adapted from [@StaPLR4]}\label{fig:staplr-flow}
\end{figure}

We denote the views by \(X^{(v)}, v = 1 \dots V\). The outcome variable is denoted by \(y\). Learning algorithms are denoted by the letter \(A\), and classifiers by the letter \(f\). For each of the views, a trained classifier \(\hat{f}_v\) is obtained by applying the base-learning algorithm \(A_{\text{b}}\) to the \(X^{(v)}\) and the outcome \(y\). We then apply \(k\)-fold cross-validation for each of these base-classifiers to obtain a vector of estimated out-of-sample predictions which we denote by \(z^{(v)}\). We assume (and recommend) that these predictions take the form of predicted probabilities instead of hard class labels. The vectors \(z^{(v)}, v = 1 \dots V\), are concatenated column-wise (i.e., \texttt{cbind}) into the matrix \(Z\). The matrix \(Z\) is then used together with outcome \(y\) to train the meta-learning algorithm \(A_{\text{m}}\) and obtain the meta-classifier \(\hat{f}_{\text{meta}}\). The final classifier is then given by using the output of the base-level classifiers as the input for the meta-classifier. This whole process can also be represented as a flowchart, which is shown in Figure \ref{fig:staplr-flow}.

Further technical details about MVS/StaPLR can be found in (Van Loon et al. 2020). For other possible choices of the meta-learner see (Van Loon, Fokkema, Szabo, et al. 2024). Although the example discussed above is hypothetical, a real application of multi-view stacking for Alzheimer's disease classification on the basis of MRI data is described in (Van Loon et al. 2022); this paper also generalized multi-view stacking to more than two levels. For multi-view stacking with missing data see
(Van Loon, Fokkema, De Vos, et al. 2024).

\section{\texorpdfstring{Why package \texttt{mvs}?}{Why package mvs?}}\label{why-package-mvs}

The multi-view stacking methodology was introduced by Li et al. (2011), and recently popularized by Garcia-Ceja, Galván-Tejada, and Brena (2018), but no software packages implementing the methodology were made publicly available. We further investigated and extended the methodology, and found that it has many favorable properties (Van Loon et al. 2020, 2024, 2022; Van Loon, Fokkema, De Vos, et al. 2024). A small number of packages for multi-view learning were available on the Comprehensive R Archive Network (CRAN) at the time of the first release of \texttt{mvs}: \texttt{Spectrum} (John and Watson 2020), \texttt{LUCIDus} (Y. Zhao 2022) and \texttt{multiview} (Ding et al. 2023), but none of these packages included multi-view stacking. Note that earlier development versions of \texttt{mvs} were also called \texttt{multiview}, but the name has been changed since it was claimed by a different package (Ding et al. 2023) which has no relation to multi-view stacking. Development versions of \texttt{mvs} have been used, among other applications, to analyze multi-view gene expression (Van Loon et al. 2020, 2024) and multi-view magnetic resonance imaging (MRI) data (Van Loon et al. 2022; Van Loon, Fokkema, De Vos, et al. 2024).

\section{Overview of package functionality}\label{overview-of-package-functionality}

\texttt{mvs} is available on \href{https://doi.org/10.32614/CRAN.package.mvs}{CRAN} and \href{https://gitlab.com/wsvanloon/mvs}{GitLab}. Below, we briefly discuss the core functionality of the package. For examples of usage see \hyperref[using-mvs-step-by-step]{Using \texttt{mvs} step-by-step}. A detailed overview of all the function arguments can be found in the \href{https://CRAN.R-project.org/package=mvs}{package reference manual}.

\subsection{Model fitting}\label{model-fitting}

The primary application of \texttt{mvs} is to fit multi-view stacking (MVS) models. The implementation of MVS is based on an extension of the \emph{Stacked Penalized Logistic Regression} (StaPLR) algorithm (Van Loon et al. 2020). \texttt{mvs} features two main functions for fitting MVS models:

\begin{itemize}
\item
  \texttt{StaPLR} is used to fit penalized and stacked penalized regression models with up to two levels. The minimum required input is the total feature matrix (\texttt{x}), the outcome variable (\texttt{y}), and a vector denoting to which view each feature corresponds (\texttt{view\_index}). The \texttt{StaPLR} function has a few special options unique to models with only two levels.
\item
  \texttt{MVS} is used to fit multi-view stacking models with two or more levels. The minimum required input is the total feature matrix (\texttt{x}), the outcome variable (\texttt{y}), and either a vector (if there are only two levels) or a matrix of dimensions {[}\(\textit{number of features} \times (\textit{number of levels} - 1)\){]} denoting to which view each feature corresponds at each level (\texttt{views}). MVS models with more than are appropriate when the data have a hierarchical multi-view structure, that is, the features are nested in views, which are themselves nested in larger views, and so on (Van Loon et al. 2022).
\end{itemize}

For more technical arguments see the documentation included with \texttt{mvs}. The individual sub-problems are optimized using coordinate descent via the R package \texttt{glmnet} (Friedman, Hastie, and Tibshirani 2010). Users of \texttt{glmnet} will feel right at home since \texttt{mvs} uses a very similar syntax.

\subsubsection{Parallelization}\label{parallelization}

One of the main advantages of MVS is that, at each level of the hierarchy, \emph{all sub-problems are independent}. This means that these sub-problems can be calculated in parallel. \texttt{mvs} supports parallel computation through \texttt{foreach} (Microsoft and Weston 2022b), assuming a parallel back-end is registered (for more information about registering a parallel back-end in R see Weston and Calaway 2015). Parallel computation can be enabled using the function argument \texttt{parallel}.

\subsubsection{Model generalizations}\label{model-generalizations}

\begin{itemize}
\tightlist
\item
  Although originally developed for binary outcome variables, \texttt{mvs} can be used to model outcome variables with different distributions. Binomial, Gaussian and Poisson distributions are currently supported through the function argument \texttt{family}.
\item
  As of version 2.0.0, \texttt{mvs} supports the use of model relaxation (as used in, e.g., the relaxed lasso (Hastie, Tibshirani, and Tibshirani 2017)). Model relaxation can be enabled for the entire hierarchy, or only for specific levels, through the function argument \texttt{relax}. Use of model relaxation is generally only sensible if \texttt{alpha} \textgreater{} 0.
\item
  As of version 2.0.0, \texttt{mvs} supports the use of adaptive weights (as used in, e.g., the adaptive lasso (Zou 2006)). Adaptive weights can be enabled for the entire hierarchy, or only for specific levels, through the function argument \texttt{adaptive}. Adaptive weights are initialized using ridge regression as described in Van Loon, Fokkema, Szabo, et al. (2024). Use of adaptive weights is generally only sensible if \texttt{alpha} \textgreater{} 0.
\item
  As of version 2.1.0, \texttt{mvs} supports the use of random forests (Breiman 2001; Liaw and Wiener 2002a) as base or meta-learner(s).
\end{itemize}

\subsection{View importance}\label{view-importance}

In a two-level StaPLR model, the meta-level regression coefficient of each view can be used as a measure of that view's importance, since these regression coefficients are effectively on the same scale (Van Loon et al. 2020). In hierarchical StaPLR/MVS models with more than two levels this does not necessarily apply, since these coefficients may correspond to different sub-models at different levels of the hierarchy. The \emph{minority report measure} (MRM) was developed to quantify the importance of a view at any level of the hierarchy (Van Loon et al. 2022). The MRM quantifies how much the prediction of the complete stacked model changes as the view-specific prediction of view \emph{i} changes from \emph{a} (default value 0) to \emph{b} (default value 1), while the other predictions are kept constant (the recommended value being the mean of the outcome variable) (Van Loon et al. 2022). As of version 2.0.0, the MRM can be calculated using \texttt{MRM}.

\subsection{Handling missing data}\label{handling-missing-data}

In practice, it is likely that not all views are measured for all observations. When a view is missing for some observations, typical approaches are:

\begin{enumerate}
\def\labelenumi{\arabic{enumi}.}
\tightlist
\item
  Remove any observations with at least one missing value (\emph{list-wise deletion})
\item
  Replace any missing values with values calculated from the observed data (\emph{imputation})
\end{enumerate}

\begin{figure}

{\centering \includegraphics[width=0.7\linewidth]{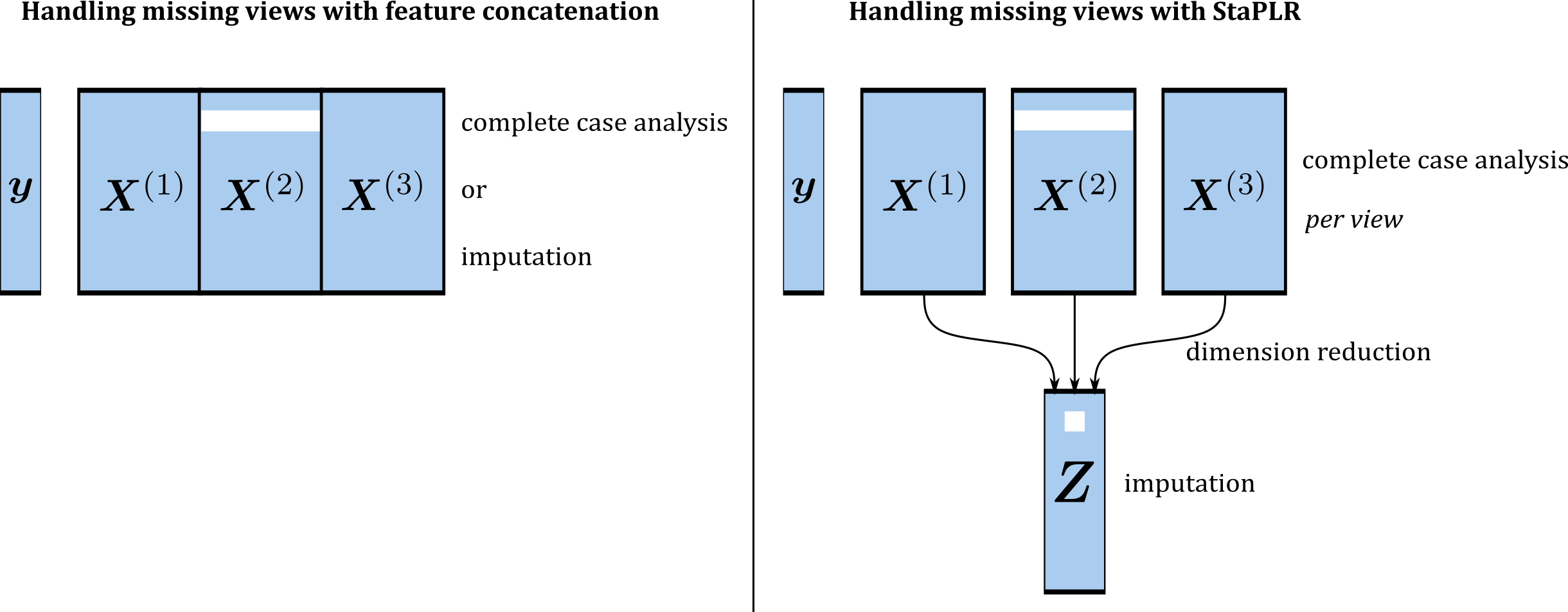} 

}

\caption{A simple graphic representation of meta-level imputation. Assume, for example, that the three views consist of, respectively, 100, 1000 and 10,000 features. Now, say that there are 10 observations which have missing values on view $X^{(2)}$. Then in traditional imputation we would have to impute 10 × 1000 = 10,000 values whereas in list-wise deletion 10 × (100 + 10,000) = 101,000 values would be deleted even though they were observed. However, in meta-level imputation only 10 values have to be imputed, and no observed values are deleted. Figure adapted from [@StaPLR4].}\label{fig:missing-data}
\end{figure}

The first approach is not recommended since it is very wasteful; often more values are removed through list-wise deletion than were initially missing. The second approach is preferable, but often unrealistic. For example, if the missing view is an MRI scan, this means having to impute millions of values, often from high-dimensional observed data, which is computationally infeasible for all but the simplest imputation algorithms (Van Loon, Fokkema, De Vos, et al. 2024). As of version 2.0.0, \texttt{mvs} therefore supports a third option: \emph{meta-level imputation}. Instead of imputing the raw data, meta-level imputation uses the complete observations for each view to generate cross-validated predictions, and performs imputation in the reduced space (see Figure \ref{fig:missing-data}).

This is much faster than traditional imputation and leads to comparable performance (Van Loon, Fokkema, De Vos, et al. 2024). It allows the use of state-of-the-art imputation algorithms which would otherwise be too computationally intensive (Van Loon, Fokkema, De Vos, et al. 2024). The following imputation methods are currently supported:

\begin{itemize}
\tightlist
\item
  \texttt{mean} performs meta-level (unconditional) mean imputation.
\item
  \texttt{mice} performs meta-level predictive mean matching. It requires the
  R package \texttt{mice} (Van Buuren and Groothuis-Oudshoorn 2011).
\item
  \texttt{missForest} performs meta-level missForest imputation. It requires
  the R package \texttt{missForest} (Stekhoven and Bühlmann 2012).
\end{itemize}

Additionally, \texttt{mvs} includes the option to `pass' the missing values through to the meta-level without imputing them, allowing the user to use a different imputation scheme of their choice. Options for missing data handling can be specified directly through model fitting using the arguments \texttt{na.action} and \texttt{na.arguments}. For more details about meta-level imputation see Van Loon, Fokkema, De Vos, et al. (2024).

\section{\texorpdfstring{Using \texttt{mvs} step-by-step}{Using mvs step-by-step}}\label{using-mvs-step-by-step}

In this section, we cover how to use the basic functions of \texttt{mvs} in practice, and show some example usage on small simulated data sets.

\subsection{Installation}\label{installation}

The current stable release can be installed directly from CRAN:

\begin{Shaded}
\begin{Highlighting}[]
\FunctionTok{install.packages}\NormalTok{(}\StringTok{"mvs"}\NormalTok{)}
\end{Highlighting}
\end{Shaded}

The current development version can be installed from GitLab using
package \textbf{\texttt{devtools}}:

\begin{Shaded}
\begin{Highlighting}[]
\NormalTok{devtools}\SpecialCharTok{::}\FunctionTok{install\_gitlab}\NormalTok{(}\StringTok{"wsvanloon/mvs@develop"}\NormalTok{)}
\end{Highlighting}
\end{Shaded}

The package can then be loaded using:

\begin{Shaded}
\begin{Highlighting}[]
\FunctionTok{library}\NormalTok{(mvs)}
\end{Highlighting}
\end{Shaded}

\subsection{Fitting a basic model}\label{fitting-a-basic-model}

We first generate a very simple simulated data set:

\begin{Shaded}
\begin{Highlighting}[]
\FunctionTok{set.seed}\NormalTok{(}\DecValTok{123}\NormalTok{)}
\NormalTok{n }\OtherTok{\textless{}{-}} \DecValTok{100}
\NormalTok{X }\OtherTok{\textless{}{-}} \FunctionTok{matrix}\NormalTok{(}\FunctionTok{rnorm}\NormalTok{(}\DecValTok{8500}\NormalTok{), }\AttributeTok{nrow=}\NormalTok{n, }\AttributeTok{ncol=}\DecValTok{85}\NormalTok{)}
\NormalTok{b }\OtherTok{\textless{}{-}} \FunctionTok{c}\NormalTok{(}\FunctionTok{rep}\NormalTok{(}\DecValTok{10}\NormalTok{, }\DecValTok{65}\NormalTok{), }\FunctionTok{rep}\NormalTok{(}\DecValTok{0}\NormalTok{, }\DecValTok{20}\NormalTok{)) }\SpecialCharTok{*}\NormalTok{ ((}\FunctionTok{rbinom}\NormalTok{(}\DecValTok{85}\NormalTok{, }\DecValTok{1}\NormalTok{, }\FloatTok{0.5}\NormalTok{)}\SpecialCharTok{*}\DecValTok{2}\NormalTok{)}\SpecialCharTok{{-}}\DecValTok{1}\NormalTok{)}
\NormalTok{eta }\OtherTok{\textless{}{-}}\NormalTok{ X }\SpecialCharTok{\%*\%}\NormalTok{ b}
\NormalTok{p }\OtherTok{\textless{}{-}} \DecValTok{1} \SpecialCharTok{/}\NormalTok{(}\DecValTok{1} \SpecialCharTok{+} \FunctionTok{exp}\NormalTok{(}\SpecialCharTok{{-}}\NormalTok{eta))}
\NormalTok{y }\OtherTok{\textless{}{-}} \FunctionTok{rbinom}\NormalTok{(n, }\DecValTok{1}\NormalTok{, p)}
\NormalTok{views }\OtherTok{\textless{}{-}} \FunctionTok{c}\NormalTok{(}\FunctionTok{rep}\NormalTok{(}\DecValTok{1}\NormalTok{,}\DecValTok{45}\NormalTok{), }\FunctionTok{rep}\NormalTok{(}\DecValTok{2}\NormalTok{,}\DecValTok{20}\NormalTok{), }\FunctionTok{rep}\NormalTok{(}\DecValTok{3}\NormalTok{,}\DecValTok{20}\NormalTok{))}
\end{Highlighting}
\end{Shaded}

This data set consists of 100 observations of 3 views. View 1 contains 45 features whereas View 2 and View 3 contain 20 features each. The first two views are truly related to the outcome, whereas View 3 is just noise. Now we will apply a simple 2-level MVS model to the data, like so:

\begin{Shaded}
\begin{Highlighting}[]
\NormalTok{fit }\OtherTok{\textless{}{-}} \FunctionTok{MVS}\NormalTok{(}\AttributeTok{x=}\NormalTok{X, }\AttributeTok{y=}\NormalTok{y, }\AttributeTok{views=}\NormalTok{views, }\AttributeTok{alphas=}\FunctionTok{c}\NormalTok{(}\DecValTok{0}\NormalTok{,}\DecValTok{1}\NormalTok{), }\AttributeTok{family=}\StringTok{"binomial"}\NormalTok{)}
\end{Highlighting}
\end{Shaded}

Argument \texttt{x} is the full data matrix, and \texttt{y} is the outcome variable. Argument \texttt{views} is a vector which denotes to which view each feature belongs. Argument \texttt{alphas} defines the penalty parameter for each level. Without going into much technical detail, a value of 0 means shrinkage is applied (akin to ridge regression (Hoerl and Kennard 1970; Le Cessie and Van Houwelingen 1992)), while a value of 1 means selection is applied (akin to the lasso (Tibshirani 1996)), while a value in between corresponds to the so-called `elastic net' (Zou and Hastie 2005). So, \texttt{alphas=c(0,1)} means we will select or discard complete views. Finally, we use \texttt{family="binomial"} since the outcome variable is binary. For other types of outcome variables we might use the \texttt{gaussian} or \texttt{poisson} family. Note that for the last two arguments we are using the default values, so they could have been omitted from the call like so:

\begin{Shaded}
\begin{Highlighting}[]
\NormalTok{fit }\OtherTok{\textless{}{-}} \FunctionTok{MVS}\NormalTok{(}\AttributeTok{x=}\NormalTok{X, }\AttributeTok{y=}\NormalTok{y, }\AttributeTok{views=}\NormalTok{views)}
\end{Highlighting}
\end{Shaded}

Instead of \texttt{MVS()} we could have also used \texttt{mvs()} since they refer to the same function. Since this model has only two levels, we could have also used the \texttt{StaPLR()} function to fit the model. However, we typically recommend using the \texttt{MVS()} function, since it is more general. Fitting the model will show a progress bar in the R console. Once the model has been fitted, we can extract model coefficients using \texttt{coef()}. In this case we are primarily interested in the coefficients at the second (i.e., meta) level:

\begin{Shaded}
\begin{Highlighting}[]
\FunctionTok{coef}\NormalTok{(fit)}\SpecialCharTok{$}\StringTok{\textquotesingle{}Level 2\textquotesingle{}}
\CommentTok{\#\textgreater{} [[1]]}
\CommentTok{\#\textgreater{} 4 x 1 sparse Matrix of class "dgCMatrix"}
\CommentTok{\#\textgreater{}                    s1}
\CommentTok{\#\textgreater{} (Intercept) {-}4.019245}
\CommentTok{\#\textgreater{} V1           3.739578}
\CommentTok{\#\textgreater{} V2           4.038358}
\CommentTok{\#\textgreater{} V3           .}
\end{Highlighting}
\end{Shaded}

These coefficients show that View 1 and View 2 were selected, while View 3 was discarded. Note that since all inputs to the meta-learner are on the same scale, these coefficients can be interpreted as-is without any further need for standardization. Their interpretation is the same as in any other logistic regression model (i.e., as predicted changes in the log-odds). The base-level coefficients can be viewed using \texttt{coef(fit)\$\textquotesingle{}Level\ 1\textquotesingle{}}, but since there are 85 of them, we will not print them here. The fitted model can also be used to predict the outcome for new observations. For example, if we have the following two new observations:

\begin{Shaded}
\begin{Highlighting}[]
\NormalTok{new\_X }\OtherTok{\textless{}{-}} \FunctionTok{matrix}\NormalTok{(}\FunctionTok{rnorm}\NormalTok{(}\DecValTok{2}\SpecialCharTok{*}\DecValTok{85}\NormalTok{), }\AttributeTok{nrow=}\DecValTok{2}\NormalTok{)}
\end{Highlighting}
\end{Shaded}

We can obtain their predicted outcomes using

\begin{Shaded}
\begin{Highlighting}[]
\FunctionTok{predict}\NormalTok{(fit, new\_X)}
\CommentTok{\#\textgreater{}           [,1]}
\CommentTok{\#\textgreater{} [1,] 0.2183148}
\CommentTok{\#\textgreater{} [2,] 0.5234955}
\end{Highlighting}
\end{Shaded}

Note that since we are using a probabilistic classifier, the predicted outcomes are probabilities rather than class labels. If we want class labels instead, we can use

\begin{Shaded}
\begin{Highlighting}[]
\FunctionTok{predict}\NormalTok{(fit, new\_X, }\AttributeTok{predtype=}\StringTok{"class"}\NormalTok{)}
\CommentTok{\#\textgreater{}      [,1]}
\CommentTok{\#\textgreater{} [1,] "0" }
\CommentTok{\#\textgreater{} [2,] "1"}
\end{Highlighting}
\end{Shaded}

\subsubsection{Random forests}\label{random-forests}

By default, \texttt{mvs} uses the extended StaPLR algorithm to fit the learners, which means all the sub-models are generalized linear models (GLMs). Depending on the outcome variable, we can set the \texttt{family} argument to either \texttt{gaussian}, \texttt{binomial} or \texttt{poisson}. In addition to GLMs, \texttt{mvs} also supports the use of random forests (Breiman 2001; Liaw and Wiener 2002a). To use random forests instead of GLMs, simply set the \texttt{type} argument to \texttt{RF}:

\begin{Shaded}
\begin{Highlighting}[]
\NormalTok{fit }\OtherTok{\textless{}{-}} \FunctionTok{MVS}\NormalTok{(}\AttributeTok{x=}\NormalTok{X, }\AttributeTok{y=}\NormalTok{y, }\AttributeTok{views=}\NormalTok{views, }\AttributeTok{type=}\StringTok{"RF"}\NormalTok{)}
\end{Highlighting}
\end{Shaded}

You can also mix and match random forests with GLMs. For example, to use random forests as the base-learners and nonnegative logistic lasso as the meta-learner, we can use

\begin{Shaded}
\begin{Highlighting}[]
\NormalTok{fit }\OtherTok{\textless{}{-}} \FunctionTok{MVS}\NormalTok{(}\AttributeTok{x=}\NormalTok{X, }\AttributeTok{y=}\NormalTok{y, }\AttributeTok{views=}\NormalTok{views, }\AttributeTok{type=}\FunctionTok{c}\NormalTok{(}\StringTok{"RF"}\NormalTok{, }\StringTok{"StaPLR"}\NormalTok{))}
\end{Highlighting}
\end{Shaded}

Note that if we extract the model coefficients using \texttt{coef()} we get only the coefficients of the meta-learner:

\begin{Shaded}
\begin{Highlighting}[]
\FunctionTok{coef}\NormalTok{(fit)}
\CommentTok{\#\textgreater{} $\textasciigrave{}Level 1\textasciigrave{}}
\CommentTok{\#\textgreater{} $\textasciigrave{}Level 1\textasciigrave{}[[1]]}
\CommentTok{\#\textgreater{} [1] NA}
\CommentTok{\#\textgreater{} }
\CommentTok{\#\textgreater{} $\textasciigrave{}Level 1\textasciigrave{}[[2]]}
\CommentTok{\#\textgreater{} [1] NA}
\CommentTok{\#\textgreater{} }
\CommentTok{\#\textgreater{} $\textasciigrave{}Level 1\textasciigrave{}[[3]]}
\CommentTok{\#\textgreater{} [1] NA}
\CommentTok{\#\textgreater{} }
\CommentTok{\#\textgreater{} }
\CommentTok{\#\textgreater{} $\textasciigrave{}Level 2\textasciigrave{}}
\CommentTok{\#\textgreater{} $\textasciigrave{}Level 2\textasciigrave{}[[1]]}
\CommentTok{\#\textgreater{} 4 x 1 sparse Matrix of class "dgCMatrix"}
\CommentTok{\#\textgreater{}                    s1}
\CommentTok{\#\textgreater{} (Intercept) {-}4.554313}
\CommentTok{\#\textgreater{} V1           5.122468}
\CommentTok{\#\textgreater{} V2           3.673454}
\CommentTok{\#\textgreater{} V3           .       }
\CommentTok{\#\textgreater{} }
\CommentTok{\#\textgreater{} }
\CommentTok{\#\textgreater{} attr(,"type")}
\CommentTok{\#\textgreater{} [1] "RF"     "StaPLR"}
\CommentTok{\#\textgreater{} attr(,"class")}
\CommentTok{\#\textgreater{} [1] "MVScoef"}
\end{Highlighting}
\end{Shaded}

This is because random forests do not have regression coefficients in the traditional sense. However, we can calculate feature importance measures using \texttt{importance()}:

\begin{Shaded}
\begin{Highlighting}[]
\FunctionTok{importance}\NormalTok{(fit)}
\CommentTok{\#\textgreater{} $\textasciigrave{}Level 1\textasciigrave{}}
\CommentTok{\#\textgreater{} $\textasciigrave{}Level 1\textasciigrave{}[[1]]}
\CommentTok{\#\textgreater{}    MeanDecreaseGini}
\CommentTok{\#\textgreater{} 1         0.7764228}
\CommentTok{\#\textgreater{} 2         0.9236304}
\CommentTok{\#\textgreater{} 3         0.6986672}
\CommentTok{\#\textgreater{} 4         0.7228284}
\CommentTok{\#\textgreater{} 5         0.9392745}
\CommentTok{\#\textgreater{}  [ reached \textquotesingle{}max\textquotesingle{} / getOption("max.print") {-}{-} omitted 40 rows ]}
\CommentTok{\#\textgreater{} }
\CommentTok{\#\textgreater{} $\textasciigrave{}Level 1\textasciigrave{}[[2]]}
\CommentTok{\#\textgreater{}    MeanDecreaseGini}
\CommentTok{\#\textgreater{} 1          4.638057}
\CommentTok{\#\textgreater{} 2          2.026891}
\CommentTok{\#\textgreater{} 3          2.411922}
\CommentTok{\#\textgreater{} 4          1.643252}
\CommentTok{\#\textgreater{} 5          1.684561}
\CommentTok{\#\textgreater{}  [ reached \textquotesingle{}max\textquotesingle{} / getOption("max.print") {-}{-} omitted 15 rows ]}
\CommentTok{\#\textgreater{} }
\CommentTok{\#\textgreater{} $\textasciigrave{}Level 1\textasciigrave{}[[3]]}
\CommentTok{\#\textgreater{}    MeanDecreaseGini}
\CommentTok{\#\textgreater{} 1          2.071363}
\CommentTok{\#\textgreater{} 2          2.067622}
\CommentTok{\#\textgreater{} 3          2.085501}
\CommentTok{\#\textgreater{} 4          2.094666}
\CommentTok{\#\textgreater{} 5          1.980755}
\CommentTok{\#\textgreater{}  [ reached \textquotesingle{}max\textquotesingle{} / getOption("max.print") {-}{-} omitted 15 rows ]}
\CommentTok{\#\textgreater{} }
\CommentTok{\#\textgreater{} }
\CommentTok{\#\textgreater{} $\textasciigrave{}Level 2\textasciigrave{}}
\CommentTok{\#\textgreater{} $\textasciigrave{}Level 2\textasciigrave{}[[1]]}
\CommentTok{\#\textgreater{} [1] NA}
\CommentTok{\#\textgreater{} }
\CommentTok{\#\textgreater{} }
\CommentTok{\#\textgreater{} attr(,"type")}
\CommentTok{\#\textgreater{} [1] "RF"     "StaPLR"}
\CommentTok{\#\textgreater{} attr(,"class")}
\CommentTok{\#\textgreater{} [1] "MVSimportance"}
\end{Highlighting}
\end{Shaded}

For an overview of the different feature importance measures available, see the \href{https://CRAN.R-project.org/package=randomForest}{\texttt{randomForest} package manual} (Liaw and Wiener 2002b). For more information about random forests in general see, for example, Breiman (2001).

\subsection{Parallel computing}\label{parallel-computing}

Multi-view stacking is computationally attractive because at any level of the model all sub-problems are independent, which means they can be computed in parallel. \texttt{mvs} supports parallel computing using \texttt{foreach} (Microsoft and Weston 2022b) and \texttt{doParallel} (Microsoft and Weston 2022a). Enabling parallel computing consists of two steps:

\begin{enumerate}
\def\labelenumi{\arabic{enumi}.}
\tightlist
\item
  Register a parallel back-end.
\item
  Use the \texttt{mvs} option \texttt{parallel\ =\ TRUE}.
\end{enumerate}

Registering a parallel back-end on a local machine is typically as simple as:

\begin{Shaded}
\begin{Highlighting}[]
\FunctionTok{library}\NormalTok{(doParallel)}
\FunctionTok{registerDoParallel}\NormalTok{(}\AttributeTok{cores =} \FunctionTok{detectCores}\NormalTok{())}
\end{Highlighting}
\end{Shaded}

However, the specifics may vary from system to system. We therefore recommend checking the \href{https://CRAN.R-project.org/package=doParallel}{doParallel vignette}. Once the parallel back-end has been registered, fitting a model using parallel computation is as simple as:

\begin{Shaded}
\begin{Highlighting}[]
\NormalTok{fit }\OtherTok{\textless{}{-}} \FunctionTok{MVS}\NormalTok{(}\AttributeTok{x=}\NormalTok{X, }\AttributeTok{y=}\NormalTok{y, }\AttributeTok{views=}\NormalTok{views, }\AttributeTok{parallel=}\ConstantTok{TRUE}\NormalTok{)}
\end{Highlighting}
\end{Shaded}

\subsection{Fitting a model with more than two levels}\label{fitting-a-model-with-more-than-two-levels}

In practice, it is possible that the multi-view structure consists of more than two levels. For example, one might have features (base level) which are grouped by brain area (middle level) and further grouped by the type of MRI scan they were obtained from (top level). Such a hierarchical analysis is described in detail in (Van Loon et al. 2022). Fitting such a model using \texttt{mvs} is very simple. Consider a modified version of the example used above:

\begin{Shaded}
\begin{Highlighting}[]
\FunctionTok{set.seed}\NormalTok{(}\DecValTok{123}\NormalTok{)}
\NormalTok{n }\OtherTok{\textless{}{-}} \DecValTok{100}
\NormalTok{X }\OtherTok{\textless{}{-}} \FunctionTok{matrix}\NormalTok{(}\FunctionTok{rnorm}\NormalTok{(}\DecValTok{8500}\NormalTok{), }\AttributeTok{nrow=}\NormalTok{n, }\AttributeTok{ncol=}\DecValTok{85}\NormalTok{)}
\NormalTok{b }\OtherTok{\textless{}{-}} \FunctionTok{c}\NormalTok{(}\FunctionTok{rep}\NormalTok{(}\DecValTok{0}\NormalTok{, }\DecValTok{15}\NormalTok{), }\FunctionTok{rep}\NormalTok{(}\DecValTok{10}\NormalTok{, }\DecValTok{40}\NormalTok{), }\FunctionTok{rep}\NormalTok{(}\DecValTok{0}\NormalTok{, }\DecValTok{30}\NormalTok{)) }\SpecialCharTok{*}\NormalTok{ ((}\FunctionTok{rbinom}\NormalTok{(}\DecValTok{85}\NormalTok{, }\DecValTok{1}\NormalTok{, }\FloatTok{0.5}\NormalTok{)}\SpecialCharTok{*}\DecValTok{2}\NormalTok{)}\SpecialCharTok{{-}}\DecValTok{1}\NormalTok{)}
\NormalTok{eta }\OtherTok{\textless{}{-}}\NormalTok{ X }\SpecialCharTok{\%*\%}\NormalTok{ b}
\NormalTok{p }\OtherTok{\textless{}{-}} \DecValTok{1} \SpecialCharTok{/}\NormalTok{(}\DecValTok{1} \SpecialCharTok{+} \FunctionTok{exp}\NormalTok{(}\SpecialCharTok{{-}}\NormalTok{eta))}
\NormalTok{y }\OtherTok{\textless{}{-}} \FunctionTok{rbinom}\NormalTok{(n, }\DecValTok{1}\NormalTok{, p)}

\NormalTok{sub\_views }\OtherTok{\textless{}{-}} \FunctionTok{c}\NormalTok{(}\FunctionTok{rep}\NormalTok{(}\DecValTok{1}\SpecialCharTok{:}\DecValTok{3}\NormalTok{, }\AttributeTok{each=}\DecValTok{15}\NormalTok{), }\FunctionTok{rep}\NormalTok{(}\DecValTok{4}\SpecialCharTok{:}\DecValTok{5}\NormalTok{, }\AttributeTok{each=}\DecValTok{10}\NormalTok{), }\FunctionTok{rep}\NormalTok{(}\DecValTok{6}\SpecialCharTok{:}\DecValTok{9}\NormalTok{, }\AttributeTok{each=}\DecValTok{5}\NormalTok{))}
\NormalTok{top\_views }\OtherTok{\textless{}{-}} \FunctionTok{c}\NormalTok{(}\FunctionTok{rep}\NormalTok{(}\DecValTok{1}\NormalTok{,}\DecValTok{45}\NormalTok{), }\FunctionTok{rep}\NormalTok{(}\DecValTok{2}\NormalTok{,}\DecValTok{20}\NormalTok{), }\FunctionTok{rep}\NormalTok{(}\DecValTok{3}\NormalTok{,}\DecValTok{20}\NormalTok{))}
\end{Highlighting}
\end{Shaded}

Here, we again have 3 views, but they are now further divided into sub-views. View 1 is divided into 3 sub-views of 15 features each, of which only the second and third sub-view are truly related to the outcome. View 2 is divided into 2 sub-views of 10 features each, of which only the first sub-view is related to the outcome. View 3 is subdivided into 4 sub-views of 4 features each, none of which are related to the outcome. The main difference when applying \texttt{MVS} to this data compared with the 2-level model is that \texttt{views} should now be a matrix where each column is a vector denoting to which view each feature corresponds \emph{at that level}. The structure is ``bottom-up'' from left to right, meaning the first column corresponds to the lowest level in the hierarchy, the second column to the level above that, and so on. For the example data above, it looks like this:

\begin{Shaded}
\begin{Highlighting}[]
\NormalTok{views }\OtherTok{\textless{}{-}} \FunctionTok{cbind}\NormalTok{(sub\_views, top\_views)}
\end{Highlighting}
\end{Shaded}

Note that although \texttt{views} has two columns, there are three levels in total: (1) the features, (2) the sub-views, and (3) the top level views. The number of levels can be determined using the \texttt{levels} argument. For each level, we need to specify the desired penalty parameter, which is indicated using the same \texttt{alphas} argument described in the previous section, except it is now a vector of length 3 instead of length 2. We will assume here that the goal is to select top level views and sub-views, but not individual features within sub-views. Finally, we also need to indicate for each level if we want to include nonnegativity constraints using argument \texttt{nnc}, which is a vector which takes value 1 if nonnegativity constraints should be applied, and 0 otherwise. We generally recommend to apply nonnegativity constraints at all levels above the feature level. The call to fit a three-level MVS model is then:

\begin{Shaded}
\begin{Highlighting}[]
\NormalTok{fit }\OtherTok{\textless{}{-}} \FunctionTok{MVS}\NormalTok{(}\AttributeTok{x=}\NormalTok{X, }\AttributeTok{y=}\NormalTok{y, }\AttributeTok{views=}\NormalTok{views, }\AttributeTok{levels=}\DecValTok{3}\NormalTok{, }\AttributeTok{alphas=}\FunctionTok{c}\NormalTok{(}\DecValTok{0}\NormalTok{,}\DecValTok{1}\NormalTok{,}\DecValTok{1}\NormalTok{), }\AttributeTok{nnc=}\FunctionTok{c}\NormalTok{(}\DecValTok{0}\NormalTok{,}\DecValTok{1}\NormalTok{,}\DecValTok{1}\NormalTok{))}
\end{Highlighting}
\end{Shaded}

The top level view coefficients are:

\begin{Shaded}
\begin{Highlighting}[]
\FunctionTok{coef}\NormalTok{(fit)}\SpecialCharTok{$}\StringTok{\textquotesingle{}Level 3\textquotesingle{}}
\CommentTok{\#\textgreater{} [[1]]}
\CommentTok{\#\textgreater{} 4 x 1 sparse Matrix of class "dgCMatrix"}
\CommentTok{\#\textgreater{}                    s1}
\CommentTok{\#\textgreater{} (Intercept) {-}3.083045}
\CommentTok{\#\textgreater{} V1           3.624991}
\CommentTok{\#\textgreater{} V2           2.465696}
\CommentTok{\#\textgreater{} V3           .}
\end{Highlighting}
\end{Shaded}

Again, Views 1 and 2 are selected, while View 3 is discarded. The sub-view coefficients for the first two can be observed by:

\begin{Shaded}
\begin{Highlighting}[]
\FunctionTok{coef}\NormalTok{(fit)}\SpecialCharTok{$}\StringTok{\textquotesingle{}Level 2\textquotesingle{}}
\CommentTok{\#\textgreater{} [[1]]}
\CommentTok{\#\textgreater{} 4 x 1 sparse Matrix of class "dgCMatrix"}
\CommentTok{\#\textgreater{}                    s1}
\CommentTok{\#\textgreater{} (Intercept) {-}4.137114}
\CommentTok{\#\textgreater{} V1           .       }
\CommentTok{\#\textgreater{} V2           3.715694}
\CommentTok{\#\textgreater{} V3           4.203360}
\CommentTok{\#\textgreater{} }
\CommentTok{\#\textgreater{} [[2]]}
\CommentTok{\#\textgreater{} 3 x 1 sparse Matrix of class "dgCMatrix"}
\CommentTok{\#\textgreater{}                    s1}
\CommentTok{\#\textgreater{} (Intercept) {-}1.493922}
\CommentTok{\#\textgreater{} V1           3.105825}
\CommentTok{\#\textgreater{} V2           .       }
\CommentTok{\#\textgreater{} }
\CommentTok{\#\textgreater{} [[3]]}
\CommentTok{\#\textgreater{} 5 x 1 sparse Matrix of class "dgCMatrix"}
\CommentTok{\#\textgreater{}                    s1}
\CommentTok{\#\textgreater{} (Intercept) 0.1201443}
\CommentTok{\#\textgreater{} V1          0.0000000}
\CommentTok{\#\textgreater{} V2          .        }
\CommentTok{\#\textgreater{} V3          .        }
\CommentTok{\#\textgreater{} V4          .}
\end{Highlighting}
\end{Shaded}

We can observe that for View 1, the second and third sub-view were selected, while for View 2, the first sub-view was selected. Note that the coefficients corresponding to the sub-views of View 3 can be ignored, since View 2 was discarded in its entirety (although in this case, the sub-view coefficients are also all zero). Note that coefficients of sub-views that are not part of the same top level view cannot be directly compared, because they are part of different sub-models. To compare the effects of sub-views that are part of different top level views, we can employ the \emph{minority report measure} (MRM) (Van Loon et al. 2022). The MRM calculates how much the final prediction of the complete stacked model changes as the prediction obtained from a sub-view changes from \texttt{a} (default value \texttt{0}) to \texttt{b} (default value \texttt{1}), while the predictions of the other views are kept constant at \texttt{constant} (the recommended value for which is \texttt{mean(y)}). The MRM for the 9 sub-views can be calculated by:

\begin{Shaded}
\begin{Highlighting}[]
\FunctionTok{MRM}\NormalTok{(fit, }\AttributeTok{constant =} \FunctionTok{mean}\NormalTok{(y), }\AttributeTok{level=}\DecValTok{2}\NormalTok{)}
\CommentTok{\#\textgreater{} [1] 0.0000000 0.5792082 0.6096366 0.3805919 0.0000000 0.0000000 0.0000000}
\CommentTok{\#\textgreater{} [8] 0.0000000 0.0000000}
\end{Highlighting}
\end{Shaded}

The value of the MRM ranges from zero to one, with larger values indicating an increased effect size. Note that if a view was excluded from the model, the value of the MRM is zero, since it has no effect on the outcome. More details about the MRM can be found in (Van Loon et al. 2022).

\subsection{Fitting a model with missing data}\label{fitting-a-model-with-missing-data}

Consider the same simulated data set we used in \hyperref[fitting-a-basic-model]{Fitting a basic model}:

\begin{Shaded}
\begin{Highlighting}[]
\FunctionTok{set.seed}\NormalTok{(}\DecValTok{123}\NormalTok{)}
\NormalTok{n }\OtherTok{\textless{}{-}} \DecValTok{100}
\NormalTok{X }\OtherTok{\textless{}{-}} \FunctionTok{matrix}\NormalTok{(}\FunctionTok{rnorm}\NormalTok{(}\DecValTok{8500}\NormalTok{), }\AttributeTok{nrow=}\NormalTok{n, }\AttributeTok{ncol=}\DecValTok{85}\NormalTok{)}
\NormalTok{b }\OtherTok{\textless{}{-}} \FunctionTok{c}\NormalTok{(}\FunctionTok{rep}\NormalTok{(}\DecValTok{10}\NormalTok{, }\DecValTok{65}\NormalTok{), }\FunctionTok{rep}\NormalTok{(}\DecValTok{0}\NormalTok{, }\DecValTok{20}\NormalTok{)) }\SpecialCharTok{*}\NormalTok{ ((}\FunctionTok{rbinom}\NormalTok{(}\DecValTok{85}\NormalTok{, }\DecValTok{1}\NormalTok{, }\FloatTok{0.5}\NormalTok{)}\SpecialCharTok{*}\DecValTok{2}\NormalTok{)}\SpecialCharTok{{-}}\DecValTok{1}\NormalTok{)}
\NormalTok{eta }\OtherTok{\textless{}{-}}\NormalTok{ X }\SpecialCharTok{\%*\%}\NormalTok{ b}
\NormalTok{p }\OtherTok{\textless{}{-}} \DecValTok{1} \SpecialCharTok{/}\NormalTok{(}\DecValTok{1} \SpecialCharTok{+} \FunctionTok{exp}\NormalTok{(}\SpecialCharTok{{-}}\NormalTok{eta))}
\NormalTok{y }\OtherTok{\textless{}{-}} \FunctionTok{rbinom}\NormalTok{(n, }\DecValTok{1}\NormalTok{, p)}
\NormalTok{views }\OtherTok{\textless{}{-}} \FunctionTok{c}\NormalTok{(}\FunctionTok{rep}\NormalTok{(}\DecValTok{1}\NormalTok{,}\DecValTok{45}\NormalTok{), }\FunctionTok{rep}\NormalTok{(}\DecValTok{2}\NormalTok{,}\DecValTok{20}\NormalTok{), }\FunctionTok{rep}\NormalTok{(}\DecValTok{3}\NormalTok{,}\DecValTok{20}\NormalTok{))}
\end{Highlighting}
\end{Shaded}

But now, assume that half of the observations have missing values on the first view:

\begin{Shaded}
\begin{Highlighting}[]
\NormalTok{X[}\DecValTok{1}\SpecialCharTok{:}\DecValTok{50}\NormalTok{, }\DecValTok{1}\SpecialCharTok{:}\DecValTok{45}\NormalTok{] }\OtherTok{\textless{}{-}} \ConstantTok{NA}
\end{Highlighting}
\end{Shaded}

If we try to fit the same model as before, we get an error:

\begin{Shaded}
\begin{Highlighting}[]
\NormalTok{fit }\OtherTok{\textless{}{-}} \FunctionTok{MVS}\NormalTok{(}\AttributeTok{x =}\NormalTok{ X, }\AttributeTok{y =}\NormalTok{ y, }\AttributeTok{views =}\NormalTok{ views)}
\CommentTok{\#\textgreater{} Error in StaPLR(X, y, view = views, skip.meta = TRUE, skip.cv = !generate.CVs, :}
\CommentTok{\#\textgreater{} Missing values detected in x. Either remove or impute missing values,}
\CommentTok{\#\textgreater{} or choose a different na.action}
\end{Highlighting}
\end{Shaded}

This is because the default value of the function parameter \texttt{na.action} is \texttt{fail}, which causes \texttt{MVS} to stop and warn the user about the presence of missing values. The error message tells us there are three possible ways to continue, namely (1) to remove all observations with missing data, (2) to impute the missing values before running \texttt{mvs} or (3) to choose a different value for \texttt{na.action}. As discussed in \hyperref[handling-missing-data]{Handling missing data}, option (1) is very wasteful, in this case deleting half of our observations. Option (2) is preferable, but quickly becomes computationally infeasible as the number of missing values and/or features increases (Van Loon, Fokkema, De Vos, et al. 2024). However, \texttt{mvs} allows for three different types of meta-level imputation using the \texttt{na.action} argument, which is much faster but obtains similar results (see \hyperref[handling-missing-data]{Handling missing data} for more details). Here we will use meta-level predictive mean matching using \texttt{mice} (Van Buuren and Groothuis-Oudshoorn 2011). To perform the meta-level imputation using \texttt{mice}, simply use:

\begin{Shaded}
\begin{Highlighting}[]
\NormalTok{fit }\OtherTok{\textless{}{-}} \FunctionTok{MVS}\NormalTok{(}\AttributeTok{x=}\NormalTok{X, }\AttributeTok{y=}\NormalTok{y, }\AttributeTok{views=}\NormalTok{views, }\AttributeTok{na.action=}\StringTok{"mice"}\NormalTok{)}
\end{Highlighting}
\end{Shaded}

Running this will print progress on both the MVS model fitting and the imputation to the console. The meta-level coefficients can again then be obtained using

\begin{Shaded}
\begin{Highlighting}[]
\FunctionTok{coef}\NormalTok{(fit)}\SpecialCharTok{$}\StringTok{\textquotesingle{}Level 2\textquotesingle{}}
\CommentTok{\#\textgreater{} [[1]]}
\CommentTok{\#\textgreater{} 4 x 1 sparse Matrix of class "dgCMatrix"}
\CommentTok{\#\textgreater{}                    s1}
\CommentTok{\#\textgreater{} (Intercept) {-}4.945277}
\CommentTok{\#\textgreater{} V1           5.916366}
\CommentTok{\#\textgreater{} V2           4.139852}
\CommentTok{\#\textgreater{} V3           .}
\end{Highlighting}
\end{Shaded}

Information about the performed imputation are stored in the \texttt{mvs} object together with the matrix of cross-validated predictions:

\begin{Shaded}
\begin{Highlighting}[]
\FunctionTok{attributes}\NormalTok{(fit}\SpecialCharTok{$}\StringTok{\textquotesingle{}Level 1\textquotesingle{}}\SpecialCharTok{$}\NormalTok{CVs)}
\CommentTok{\#\textgreater{} $dim}
\CommentTok{\#\textgreater{} [1] 100   3}
\CommentTok{\#\textgreater{} }
\CommentTok{\#\textgreater{} $imputation\_method}
\CommentTok{\#\textgreater{} result.1 result.2 result.3        y }
\CommentTok{\#\textgreater{}    "pmm"       ""       ""       "" }
\CommentTok{\#\textgreater{} }
\CommentTok{\#\textgreater{} $number\_of\_imputations}
\CommentTok{\#\textgreater{} [1] 5}
\CommentTok{\#\textgreater{} }
\CommentTok{\#\textgreater{} $additional\_arguments\_passed\_to\_mice}
\CommentTok{\#\textgreater{} list()}
\end{Highlighting}
\end{Shaded}

This shows us that the first view was imputed using predictive mean matching (``pmm''), whereas the other views were not imputed (since they had no missing values). Note that the outcome variable \texttt{y} was also used in the imputation process, as is generally recommended (Van Buuren 2018). The attributes also show us that the given matrix of cross-validated predictions is an average of 5 different imputations. The number of imputations, or any other \texttt{mice} arguments can be changed by providing a list of arguments and their values using the \texttt{na.arguments} option. For example, to change the number of imputations to 10 use:

\begin{Shaded}
\begin{Highlighting}[]
\NormalTok{fit }\OtherTok{\textless{}{-}} \FunctionTok{MVS}\NormalTok{(}\AttributeTok{x=}\NormalTok{X, }\AttributeTok{y=}\NormalTok{y, }\AttributeTok{views=}\NormalTok{views, }\AttributeTok{na.action=}\StringTok{"mice"}\NormalTok{, }\AttributeTok{na.arguments=}\FunctionTok{list}\NormalTok{(}\AttributeTok{m =} \DecValTok{10}\NormalTok{))}
\end{Highlighting}
\end{Shaded}

\begin{Shaded}
\begin{Highlighting}[]
\FunctionTok{attributes}\NormalTok{(fit}\SpecialCharTok{$}\StringTok{\textquotesingle{}Level 1\textquotesingle{}}\SpecialCharTok{$}\NormalTok{CVs)}
\CommentTok{\#\textgreater{} $dim}
\CommentTok{\#\textgreater{} [1] 100   3}
\CommentTok{\#\textgreater{} }
\CommentTok{\#\textgreater{} $imputation\_method}
\CommentTok{\#\textgreater{} result.1 result.2 result.3        y }
\CommentTok{\#\textgreater{}    "pmm"       ""       ""       "" }
\CommentTok{\#\textgreater{} }
\CommentTok{\#\textgreater{} $number\_of\_imputations}
\CommentTok{\#\textgreater{} [1] 10}
\CommentTok{\#\textgreater{} }
\CommentTok{\#\textgreater{} $additional\_arguments\_passed\_to\_mice}
\CommentTok{\#\textgreater{} $additional\_arguments\_passed\_to\_mice$m}
\CommentTok{\#\textgreater{} [1] 10}
\end{Highlighting}
\end{Shaded}

In addition to imputation using \texttt{mice}, meta-level imputation using the \texttt{mean} and meta-level imputation with \texttt{missForest} (Stekhoven and Bühlmann 2012) are also supported. Note that there is another possible value for \texttt{na.action}, namely \texttt{pass}. Using this value does not perform any imputation, but instead ``passes'' the missingness onto the meta-level:

\begin{Shaded}
\begin{Highlighting}[]
\NormalTok{fit }\OtherTok{\textless{}{-}} \FunctionTok{MVS}\NormalTok{(}\AttributeTok{x=}\NormalTok{X, }\AttributeTok{y=}\NormalTok{y, }\AttributeTok{views=}\NormalTok{views, }\AttributeTok{na.action=}\StringTok{"pass"}\NormalTok{)}
\end{Highlighting}
\end{Shaded}

\begin{Shaded}
\begin{Highlighting}[]
\NormalTok{fit}\SpecialCharTok{$}\StringTok{\textasciigrave{}}\AttributeTok{Level 1}\StringTok{\textasciigrave{}}\SpecialCharTok{$}\NormalTok{CVs}
\CommentTok{\#\textgreater{}              [,1]       [,2]      [,3]}
\CommentTok{\#\textgreater{}   [1,]         NA 0.84404597 0.5555556}
\CommentTok{\#\textgreater{}   [2,]         NA 0.55530209 0.5555556}
\CommentTok{\#\textgreater{}   [3,]         NA 0.56042449 0.5101729}
\CommentTok{\#\textgreater{}   [4,]         NA 0.72310130 0.5787974}
\CommentTok{\#\textgreater{}   [5,]         NA 0.76159335 0.5017087}
\CommentTok{\#\textgreater{}   [6,]         NA 0.47206796 0.5117947}
\CommentTok{\#\textgreater{}   [7,]         NA 0.73552713 0.5000000}
\CommentTok{\#\textgreater{}   [8,]         NA 0.39004445 0.5504475}
\CommentTok{\#\textgreater{}   [9,]         NA 0.82898606 0.5475928}
\CommentTok{\#\textgreater{}  [10,]         NA 0.26547150 0.5502753}
\CommentTok{\#\textgreater{}  [ reached \textquotesingle{}max\textquotesingle{} / getOption("max.print") {-}{-} omitted 90 rows ]}
\end{Highlighting}
\end{Shaded}

This option is primarily useful for implementing custom imputation schemes other than those supported by \texttt{mice} or \texttt{missForest}.

\section{Acknowledgements}\label{acknowledgements}

I acknowledge the contribution of Marjolein Fokkema, who helped prepare \texttt{mvs} for submission to CRAN and worked on the implementation of model relaxation and random forests. I acknowledge the support of Mark de Rooij and Botond Szabo during the earlier stages of this project.

\section*{References}\label{references}
\addcontentsline{toc}{section}{References}

\phantomsection\label{refs}

\end{document}